\documentclass[useAMS,usenatbib]{mn2e}
\usepackage{epsfig}
\usepackage{amssymb}
\usepackage{amsmath}
\usepackage{lscape}
\usepackage{longtable}
\usepackage[normal]{caption}
\usepackage{appendix}
\usepackage{times}

\def\arcsec{$^{\prime\prime}$}

\title[\textit{Herschel}/PACS Observations of the Host Galaxy of GRB\,031203]{\textit{Herschel}/PACS Observations of the Host Galaxy of GRB\,031203 \thanks{{\it Herschel} is an ESA space observatory with science instruments provided by European-led Principal Investigator consortia and with important participation from NASA.}}

\author[M.~Symeonidis et al.]
{\parbox{\textwidth}{\raggedright M.~Symeonidis,$^{1,2}$\thanks{E-mail: \texttt{m.symeonidis@sussex.ac.uk}}
S. R. ~Oates,$^{2}$
M. ~de Pasquale,$^{2}$
M. J. ~Page,$^{2}$
K. Wiersema,$^{3}$
R. Starling,$^{3}$
P. Schady,$^{4}$
N. Seymour$^{5}$
and B. O'Halloran$^{6}$}\vspace{0.4cm} \\
\parbox{\textwidth}{\raggedright $^{1}$Astronomy Centre, Dept. of Physics \& Astronomy, University of Sussex, Brighton BN1 9QH, UK\\
$^{2}$Mullard Space Science
  Laboratory, University College London, Holmbury St. Mary, Dorking,
  Surrey RH5 6NT, UK\\
$^{3}$Department of Physics $\&$ Astronomy, University of Leicester, University Road, Leicester, LE1 7RH, United Kingdom\\
$^{4}$Max-Planck-Institut für extraterrestrische Physik,
Giessenbachstrasse, 85748, Garching, Germany\\
$^{5}$CSIRO Astronomy \& Space Science, PO Box 76, Epping, NSW 1710,
Australia\\
$^{6}$Imperial College London, Astrophysics, Blackett Laboratory, Prince Consort Road, London SW7 2AZ, UK }}

\begin{document}

\date{Accepted  Received; in original form}

\pagerange{\pageref{firstpage}--\pageref{lastpage}} \pubyear{2013}

\maketitle

\label{firstpage}

\begin{abstract}
We present \textit{Herschel}/PACS observations of the nearby ($z=0.1055$) dwarf
galaxy that has hosted the long gamma ray burst (LGRB) 031203. Using the PACS data we have been able to place constraints on
the dust temperature, dust mass, total infrared luminosity and infrared-derived
star-formation rate ($SFR$) for this object. We find that the GRB host galaxy (GRBH)
031203 has a total infrared luminosity of 3$\times$10$^{10}$\,L$_{\odot}$ placing
it in the regime of the IR-luminous galaxy population. Its dust
temperature and specific $SFR$ are comparable to that of many high-redshift ($z$=0.3-2.5)
infrared (IR)-detected GRB hosts ($T_{\rm dust}>$40\,K;
$sSFR>$10\,Gyr$^{-1}$), however its dust-to-stellar mass ratio is lower than
what is commonly seen in IR-luminous galaxies. Our results
suggest that GRBH\,031203 is undergoing a strong starburst episode and
its dust properties are different to those of local dwarf
galaxies within the same metallicity and stellar mass
range. Furthermore, our measurements place it in a distinct class to the well
studied nearby host of GRB\,980425 ($z$=0.0085), confirming the
notion that GRB host galaxies can span a large range in properties even at similar cosmological epochs,
making LGRBs an ideal tool in selecting
samples of star-forming galaxies up to high redshift.

\end{abstract}


\section{Introduction}
The currently favoured scenario for 
the origin of long gamma-ray bursts (LGRBs) is that they result from the collapse of 
massive, metal-poor stars and as a
result LGRBs are thought to mark the sites of cosmic star-formation (e.g. Christensen et al. 2004\nocite{CHG04}; Tanvir et
al. 2004\nocite{Tanvir04}). In comparison to
the typical flux limited galaxy surveys, LGRBs are considered more unbiased
identifiers of star-forming galaxies because their occurrence and detection
is independent of many galaxy properties, such as extinction (e.g. Ramirez-Ruiz et al. 2002\nocite{RRTB02}; Watson et al. 2011\nocite{Watson11}). As a result, the follow up of
LGRB hosts plays a key role in determining the cosmic star-formation
history and at high redshift ($z>$5) it has important implications for our understanding 
of the epoch of re-ionisation (e.g Robertson $\&$ Ellis 2012\nocite{RE12}).

\begin{figure*}
\begin{tabular}{c|c|c|c|c|c}
\epsfig{file=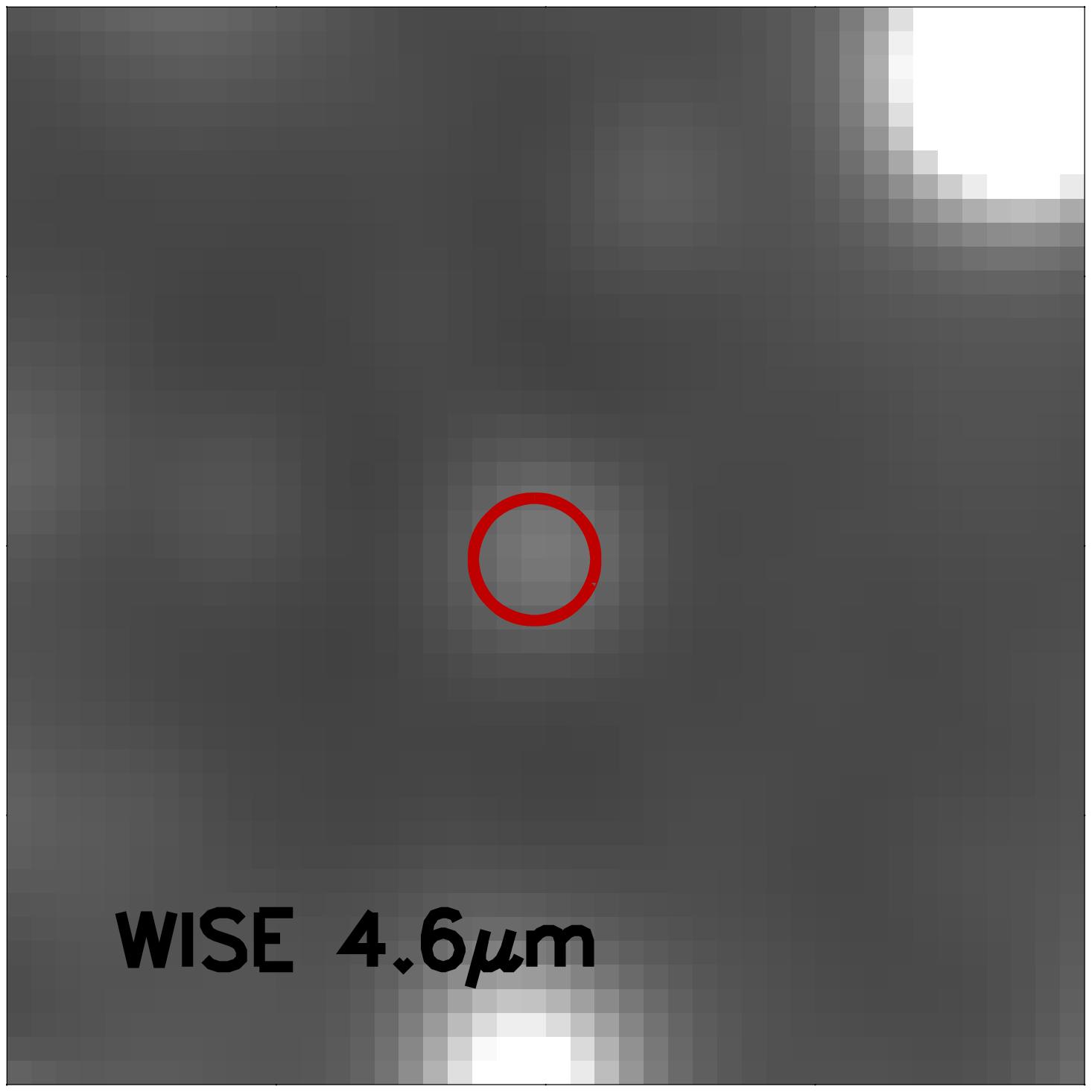,width=0.14\linewidth} &\epsfig{file=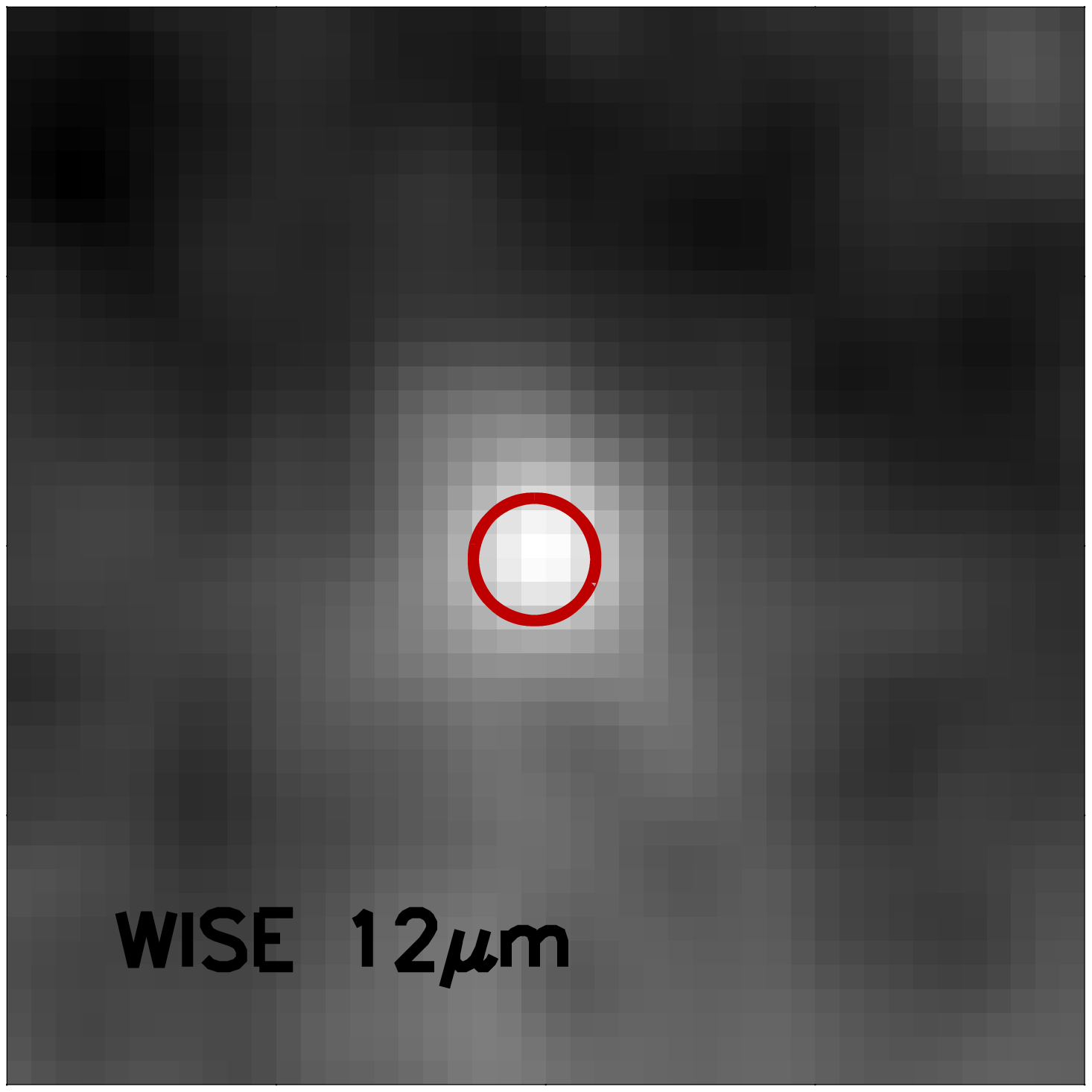, width=0.14\linewidth} &\epsfig{file=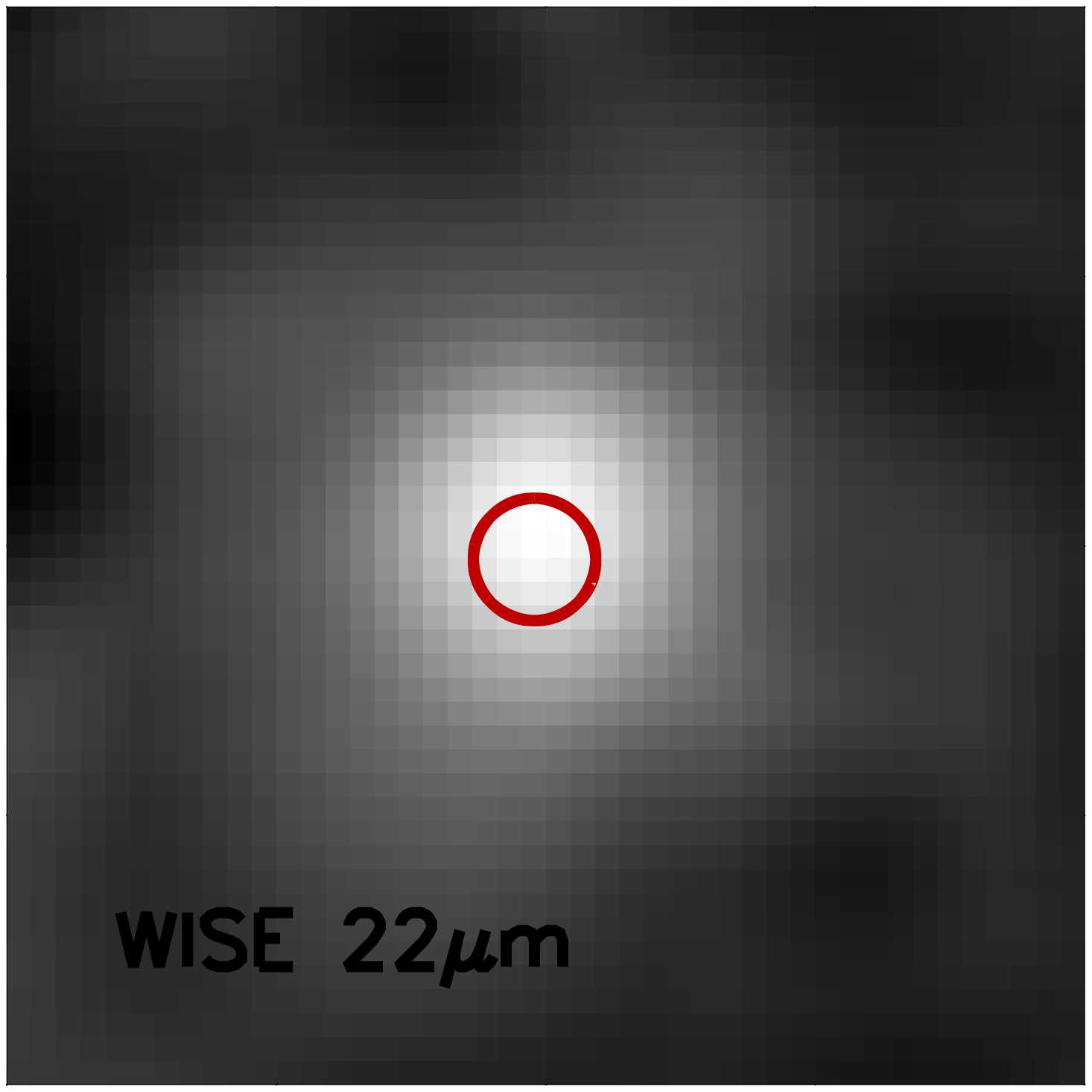, width=0.14\linewidth} &\epsfig{file=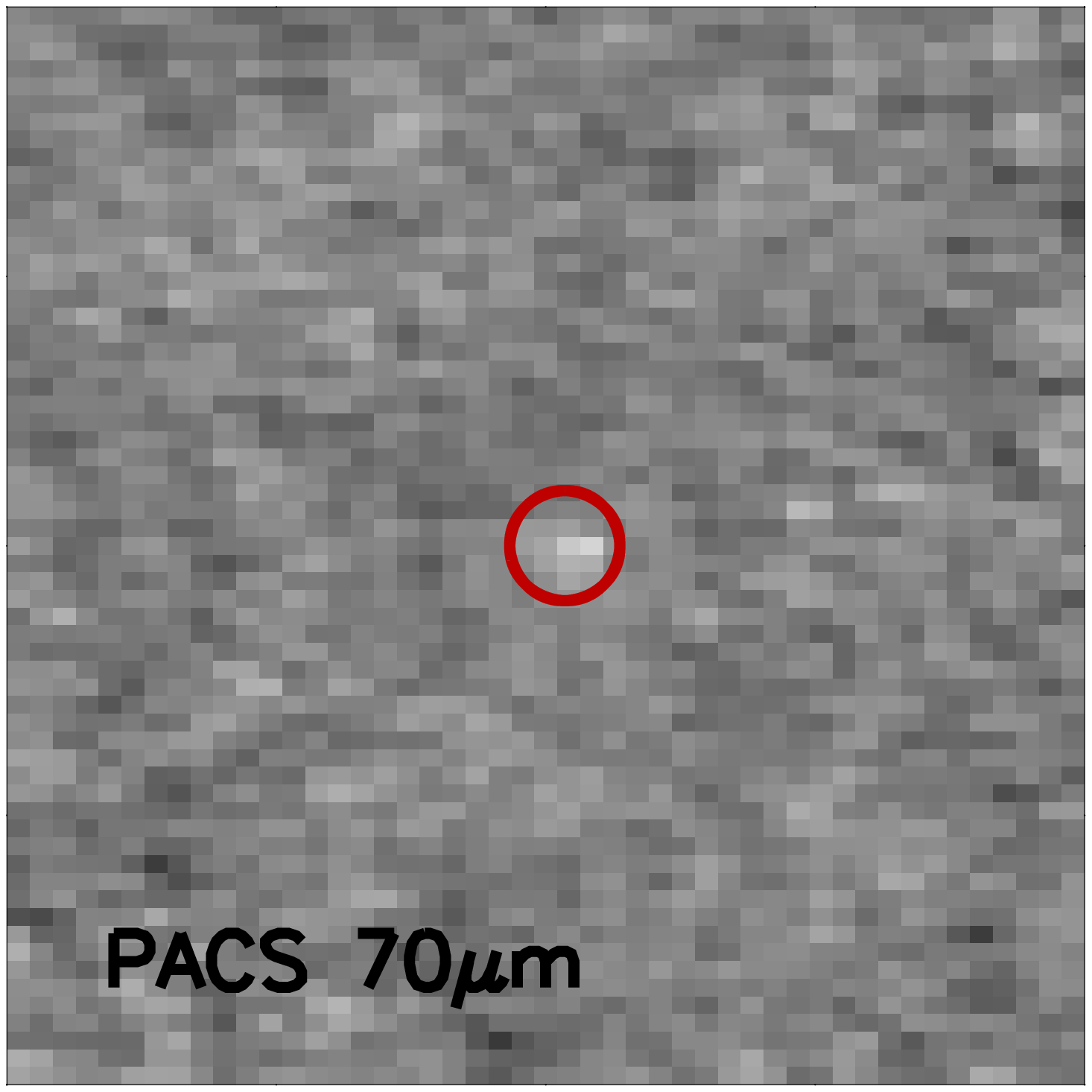, width=0.14\linewidth} &\epsfig{file=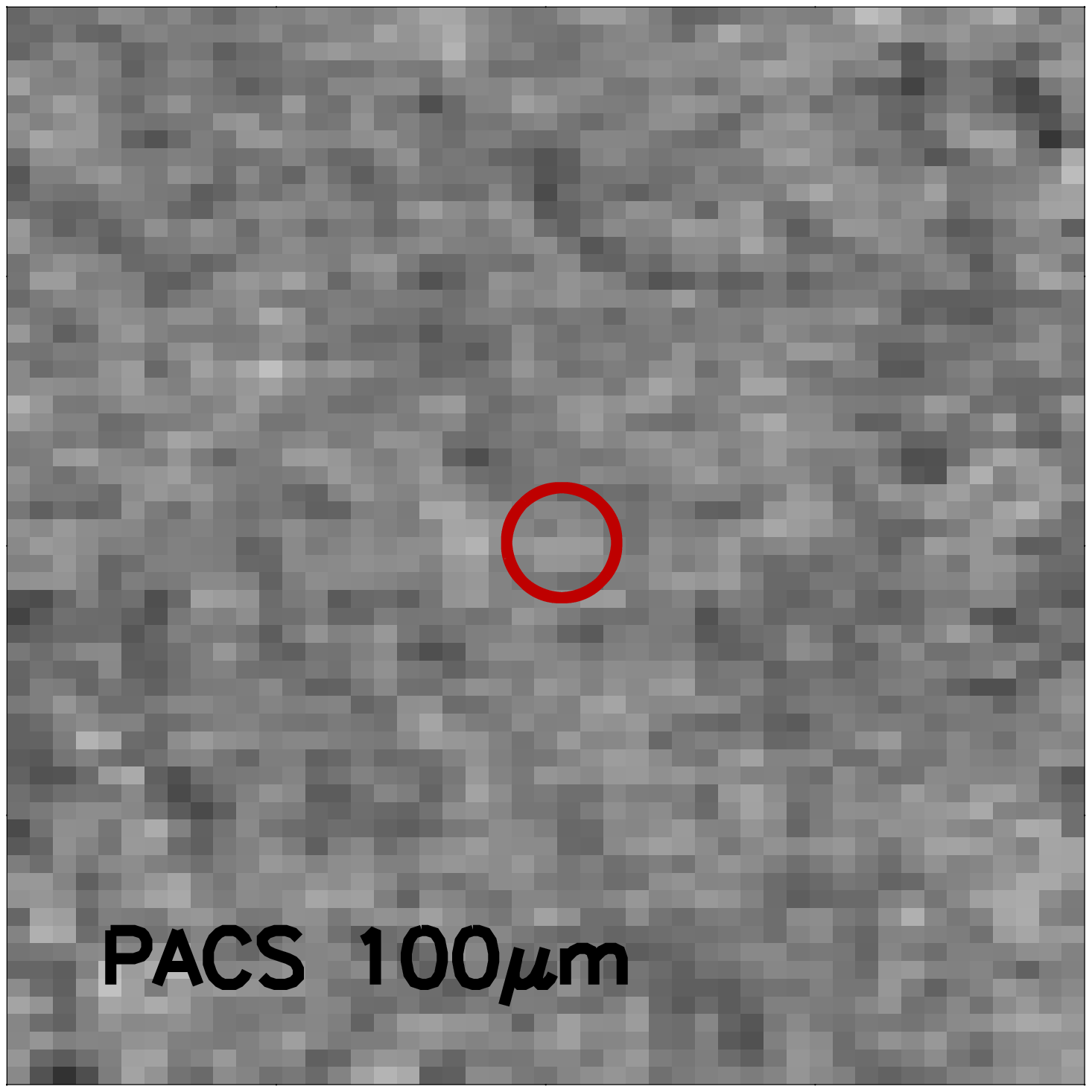, width=0.14\linewidth} &\epsfig{file=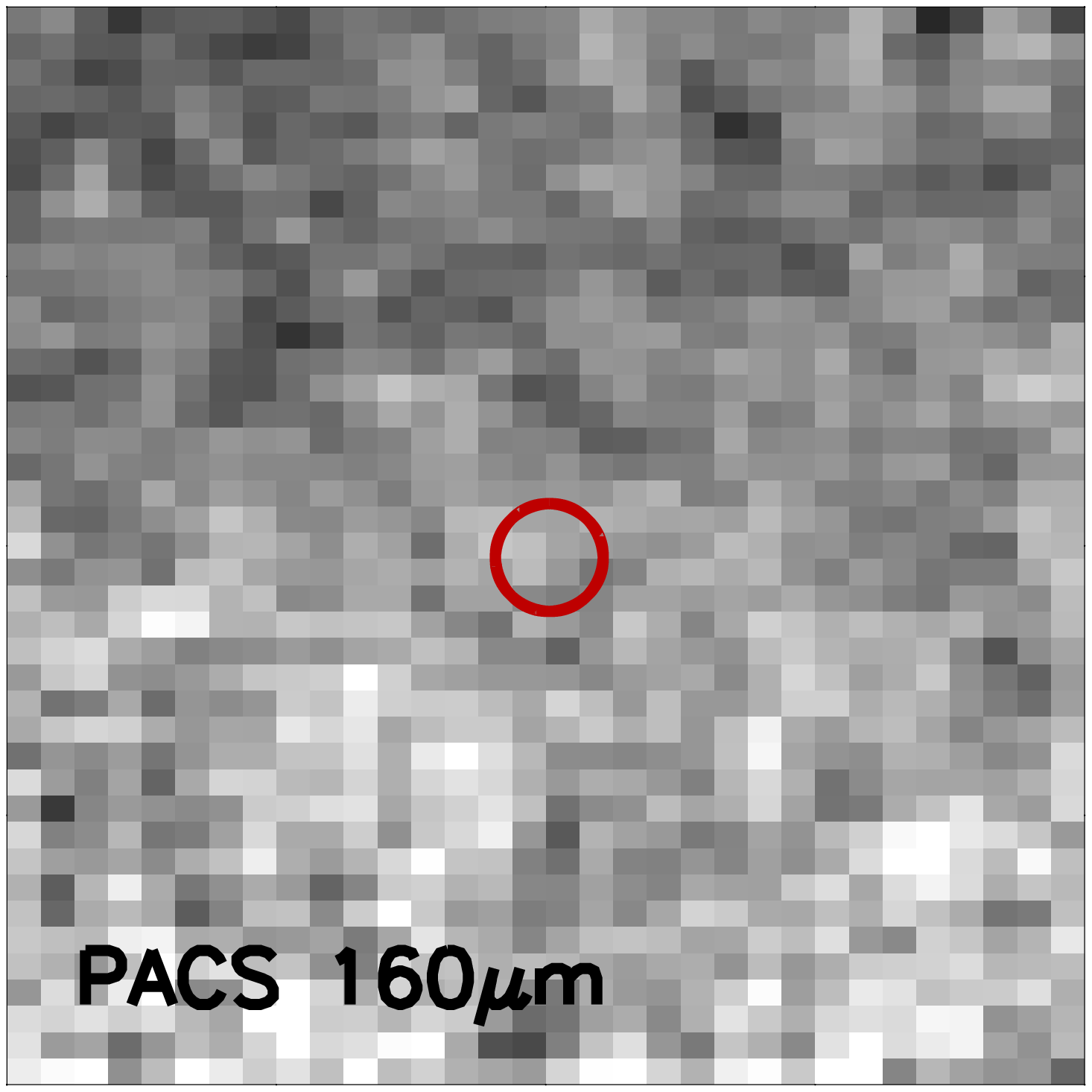, width=0.14\linewidth} \\
\end{tabular}
\caption{\textit{WISE} and PACS images of GRBH\,031203 (RA
  08:02:30.18; DEC -39:51:03.52). North is up and East is left. Each
  cut-out is 1 arcmin by 1 arcmin. The red circle denotes the position
of the galaxy and has a radius of 6 arcseconds. } 
\label{fig:images}
\end{figure*}

Targeted observations of LGRB host galaxies have shown them to be
young, spanning a large range of properties (e.g. Kr{\"u}hler et al. 2011\nocite{Kuhler11}; Perley et
al. 2013\nocite{Perley13}), however, the role of dust in the star
formation history, chemical enrichment and evolution of these sources
is still largely unclear.
Determining host galaxy properties, such as dust extinction, using the LGRB afterglows
(e.g. Schady et al. 2007\nocite{Schady07}; Starling et
al. 2007\nocite{Starling07}) target only line-of-sight conditions and
hence may only be
typical of the local environment around the GRB event rather
than the host itself, particularly since there is evidence that
measurements over small scales are not necessarily representative of
global galaxy attributes (e.g. Padoan et
al. 2006\nocite{Padoan06}). Furthermore, as the abundance of dust in star-forming
galaxies implies that they radiate a large
fraction of their energy in the infrared (IR), star-formation rates (SFRs) derived from optical observations cannot account for deeply dust-embedded star-forming
regions and have to rely on uncertain extinction corrections.  
Examining the infrared part of the spectral energy distribution (SED)
is thus essential in order to
determine the total energy budget and global SFR, as well as dust properties such as
temperature, mass and extinction. Indeed, estimating the
gas-to-dust mass ratio, a key ingredient in constraining the chemical
evolution of galaxies, is not feasible without infrared
observations. 
Ultimately, determining the role of dust in young environments, such as those found in LGRB host
galaxies, is crucial for our understanding of primordial galaxy
formation in the early Universe. 

To date, out of the GRB host galaxies targetted in the infrared with \textit{Spitzer}/MIPS and JCMT/SCUBA, only a small fraction have been detected (e.g. Tanvir et al. 2004\nocite{Tanvir04}; Le
Floc'h et al. 2006\nocite{LeFloch06}; Micha{\l}owski et al. 2008) and
more recently with \textit{Herschel} (Micha{\l}owski et al. 2013; Hunt et al. 2014). 
Here we present \textit{Herschel} (Pilbratt et
al. 2010\nocite{Pilbratt10})/PACS (Poglitsch et al. 2010\nocite{Poglitsch10}) observations of the host galaxy
of GRB\,031203 (hereafter referred to as GRBH\,031203). GRBH\,031203 is a metal poor compact dwarf at $z=$0.1055 (Prochaska et
al. 2004\nocite{Prochaska04}, Margutti et al. 2007\nocite{Margutti07}; Levesque
et al. 2010\nocite{Levesque10}; Han et al. 2010\nocite{Han10}), one of
the 4 local ($z \lesssim 0.1$) galaxies that
have hosted a GRB event. 
Our aim is to place constraints on the dust properties and IR SED of this galaxy,
which has not been possible to date due to lack of far-IR observations. 
This letter is laid out as follows: Section \ref{sec:data} outlines the
data used, section \ref{sec:measurements} presents our SED measurements
and in section \ref{sec:properties} we discuss the properties of
GRBH\,031203. Our summary and conclusions can
be found in Section \ref{sec:conclusions}. Throughout we adopt a concordance cosmology of H$_0$=70\,km\,s$^{-1}$Mpc$^{-1}$, $\Omega_{\rm M}$=1-$\Omega_{\rm \Lambda}$=0.3.

\section{Data}
\label{sec:data}

\subsection{Herschel Data}
We acquired \textit{Herschel}/PACS 70, 100 and 160$\mu$m data as
part of the OT2 GO proposal cycle in 2013
(P.I. Symeonidis). Reduction of the PACS images was performed with the Herschel Interactive Processing Environment (HIPE) version 10.3.0.

From the PANIC J-band image in Gal-Yam et al. (2004\nocite{GalYam04}) we see that
GRBH\,031203 is a compact source ($\sim$\,2\arcsec). On the position of
the host galaxy (see Fig. \ref{fig:images}) we perform aperture photometry with the aperture photometry tool
(APT\footnote{http://www.aperturephotometry.org/}). We use an aperture
radius of 12\arcsec and the prescribed aperture correction of 0.794 to
determine the 70\,$\mu$m flux density. For the 100\,$\mu$m flux
density, the same aperture size and an aperture correction of 0.766 is
used, whereas for the 160\,$\mu$m flux density an aperture size of
22\arcsec and an aperture correction of 0.810 is used. 
To calculate 1$\sigma$ photometric uncertainties and upper
limits we use the
method outlined in the PACS documents (Muller et al. 2011 and Balogh et
al. 2013\footnote{http://herschel.esac.esa.int/twiki/bin/view/Public/
PacsCalibrationWeb?template=viewprint}). We
check the validity of this method by calculating the standard deviation of the distribution of flux densities measured in 20 same
-sized apertures placed on empty parts of the map near the source. We
find the 1$\sigma$ deviation to be the consistent with both methods. 
The PACS photometry is displayed in Table \ref{table:data}.

\subsection{Multiwavelength data}
We take optical photometry from Margutti et al. (2007\nocite{Margutti07}) at
$\lambda<$800\,nm, from Cobb et al. (2004\nocite{Cobb04}) and Prochaska et al. (2004\nocite{Prochaska04})
at $\lambda<$3\,$\mu$m. We also use the IRAC and MIPS 24\,$\mu$m
photometry from Watson et
al. (2011\nocite{Watson11}). Watson et al. (2011) report that this source was also observed in the MIPS SED mode, which provides low
resolution spectroscopy between 52 to 100 $\mu$m, however it was not
detected and has an upper limit of 40\,mJy. Our
estimated flux density at 70\,$\mu$m is consistent with this value. 
Finally, we also retrieve near and mid-IR data from the Wide-field
Infrared Survey Explorer survey (\textit{WISE}; Wright et al.
2010\nocite{Wright10}) using the newly updated
AllWISE Source Catalogue\footnote{http://wise2.ipac.caltech.edu/docs/release/allwise/}. Table \ref{table:data} shows the \textit{WISE} photometry for
GRBH\,031203.

\begin{table*}
\centering
\caption{The \textit{WISE} (3.6, 5.4, 12 and 22\,$\mu$m) and \textit{Herschel} PACS (70, 100 and 160\,$\mu$m) flux
  density measurements (and
  1$\sigma$ errors) in mJy. The 3$\sigma$ upper limits for 100 and 160\,$\mu$m are shown in brackets.}
\begin{tabular}{c|c|c|c|c|c|c|}
\hline 
\textit{WISE} 3.6$\mu$m & \textit{WISE} 5.4$\mu$m & \textit{WISE} 12\,$\mu$m& \textit{WISE} 22$\mu$m &PACS 70\,$\mu$m & PACS 100$\mu$m & PACS 160$\mu$m\\
\hline
0.11$\pm$0.005 &   0.086$\pm$0.009   &   1.74$\pm$0.12   &   11.3$\pm$1.0&39$\pm$18&22$\pm$16 ($<$70) &47$\pm$34 ($<$149)\\
\hline
\end{tabular}
\label{table:data}
\end{table*}

\section{SED measurements}
\label{sec:measurements}
\begin{figure}
\epsfig{file=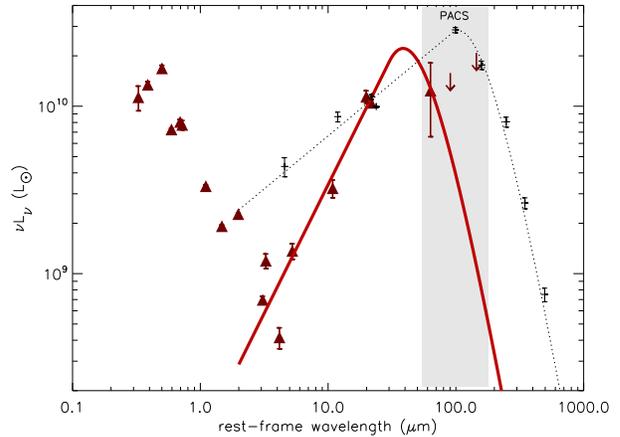, width=0.99\linewidth} 
\caption{The SED of the GRBH\,031203: photometry (red triangles and
  upper limits) and best-fit SED model (red curve). For
  comparison we overplot the SED of GRBH\,980425 normalised at 20$\mu$m to the SED of GRBH\,031203 (dotted line: best fit
  model; black points: photometry from Micha{\l}owski et
  al. 2013).}
\label{fig:SED}
\end{figure}

We fit the photometry of GRBH\,031203, with a simple SED model, which
combines the grey-body function (GBF) for the far-IR and a power-law
(PL) for the
mid-IR at a critical frequency $\nu_{\ast}$ (see also Blain
et al. 2003\nocite{BBC03}; Younger et al. 2009\nocite{Younger09}) as follows:
\begin{equation}
F_{\nu} \propto  \left\{
  \begin{array}{l l}
    \frac{\nu ^{3+\beta}}{e^{(h\nu/kT_{\rm dust})} - 1} & \quad $if$ \quad \nu < \nu_{\ast}\\
    \nu^{\alpha} & \quad  $if$ \quad \nu > \nu_{\ast}
  \end{array} \right.\
\end{equation}
where $F_{\nu} $ is the flux density, $h$ is the Planck constant, $c$ is the speed of light in a
vacuum, $k$ is the Boltzmann constant, $T_{\rm dust}$ is the temperature of the
grey-body function and $\beta$ is the emissivity --- we adopt $\beta$=1.5, consistent with studies of the far-IR emissivity of large
grains (Desert, Boulanger $\&$ Puget 1990\nocite{DBP90}). At the critical
frequency $\nu_{\ast}$ the slopes of the two functions are equal
and hence $\alpha$=$d$log GBF/$d$log$\nu$.
We perform  $\chi ^2$ fitting to the dust component of the galaxy SED,
i.e. from 4$\mu$m onwards, in order to obtain the normalisation, $T$ and $\alpha$. The 0.68 lower and upper confidence limits for our computed
parameters resulting from the fits (e.g. temperature, total infrared luminosity etc.) are calculated according to the
prescribed $\chi^2$ confidence intervals for one interesting
parameter, namely $\chi^2_{\rm min}+1$, where $\chi^2_{\rm min}$ is
the minimum $\chi^2$. Fig. \ref{fig:SED} shows the SED fit to the photometry.

We compute the total IR luminosity ($L_{\rm IR}$) in the 8--1000\,$\mu$m
range, $T_{\rm dust}$ (the average dust temperature of the galaxy
representing the peak of the SED) and the dust mass ($M_{\rm dust}$)
--- see Table \ref{table:measurements}. $M_{\rm dust}$ is calculated as follows:
\begin{equation}
M_{\rm dust}=\frac{f_{\nu, \rm rest} D_{\rm L}^2}{B(\nu_{\rm
    rest},T_{\rm dust, rest})
  \kappa_{\rm rest}} 
\end{equation}
where, $D_{\rm L}$ is the luminosity distance, $B(\nu_{\rm
  rest},T_{\rm dust, rest})$ is the black body function (in units of
flux density), $f_{\nu,\rm rest}=\frac{f_{\nu,\rm obs}}{(1+z)}$, $\kappa_{\rm rest}=\kappa_{850\mu m} (\frac{\nu_{\rm rest}}{\nu_{\rm
    850\mu m}})^{\beta}$ and $\kappa_{850\mu m}$=0.0431\,m$^2$\,kg$^{-1}$  taken from Li and
Draine (2001\nocite{LD01}). Here, we take $f_{\nu,\rm obs}$ as the observed flux
density at 250\,$\mu$m computed by using the model SED and
$\nu_{\rm rest}$ is the rest-frame frequency equivalent to 250\,$\mu$m
(observed). 

We convert $L_{\rm IR}$ to
SFR using the prescription of Kennicutt (1998), finding
5.06$^{+0.07}_{-0.91}$\,M$_{\odot}$/yr as the SFR of GRBH\,031203. This is consistent with SFRs estimated from radio measurements: 
Stanway et al. (2010\nocite{SDL10}) report an SFR of
4.8$^{+1.4}_{-0.9}$\,M$_{\odot}$/yr and Micha{\l}owski et al. (2012\nocite{Michalowski12})
report an SFR of 3.83$\pm$0.69\,M$_{\odot}$/yr. However, SFRs
calculated from optical and UV measurements (mainly H$\alpha$ and
[OII] lines and UV continuum) are
in disagreement varying from 0.4 to 13\,M$_{\odot}$/yr (Prochaska et
al. 2004; Margutti et al 2007; Savaglio et al. 2009\nocite{SGLB09}; Svensson et
al. 2010\nocite{Svensson10}; Levesque et al. 2010\nocite{Levesque10};
Guseva et al. 2011\nocite{Guseva11}). It is likely that
the discrepancies between the optical SFR measurements in literature are
mainly due to discrepant extinction corrections. Stanway et al. (2010)
also propose that optically-derived SFR measurements for GRBH\,031203 are subject
to AGN contamination, based on the conclusions of
Levesque et al. (2010), however, many authors dispute the presence of
an AGN (e.g. Prochaska et al. 2004; Margutti et al. 2007, Watson et
al. 2011). 

We calculate a specific star-formation rate ($sSFR$) of 20.2\,Gyr$^{-1}$, using the stellar mass of 2.5$\times$10$^8$\,M$_{\odot}$ reported in Guseva et
al. (2011\nocite{Guseva11}). The computed $sSFR$ is within the range measured in other
LGRB host galaxies (e.g. Castro Cer{\'o}n et al. 2006\nocite{Castro-Ceron06}; Savaglio et al. 2009; Perley et al. 2013).

Fig. \ref{fig:SED} shows that the SED of GRBH\,031203 is warm peaking
at $\lambda_{\rm peak}$=40$^{+19}_{-8}$\,$\mu$m, consistent with a predisposition of low
metallicity star-forming galaxies to have
SEDs which peak at short wavelengths (e.g. Galametz et al. 2009;
2011; 2013\nocite{Galametz09}\nocite{Galametz11}\nocite{Galametz13}). For comparison, we overplot the
SED of the nearest GRB host galaxy, GRBH\,980425 ($z$=0.0085), the only other local
GRB host with \textit{Herschel} observations. We fit the IR photometry
reported in Micha{\l}owski et al. (2013)\nocite{Michalowski13} with
the combination of grey-body/power-law functions as described earlier. It is interesting to note that GRBH\,980425 has a
cooler SED, which peaks at longer wavelengths ($\lambda_{\rm
peak}$=104\,$\mu$m and $T_{\rm dust}$=25\,K). 

Note that although GRBH\,031203 is not significantly detected in the
far-IR, the PACS data allows us to significantly constrain
its IR SED shape and hence dust properties, particularly when combined
with the mid-IR data from \textit{WISE} and \textit{Spitzer}/MIPS. In
fact, even a simple visual inspection of the available photometry for
GRBH\,031203 in Fig. \ref{fig:SED} indicates that the SED peak must
be between 20 and 70\,$\mu$m.

\begin{table}
\centering
\caption{The derived properties from SED fitting, including upper and
  lower 1$\sigma$ errors. }
\begin{tabular}{|c|c|c|c|c|c|}
\hline 
$\lambda_{\rm peak}$&$T_{\rm dust}$ & log\,$L_{\rm IR}$ & log\,$M_{\rm dust}$ &$SFR_{\rm IR}$ &
$sSFR_{\rm IR}$ \\
($\mu$m)& (K) & (L$_{\odot}$) & (M$_{\odot}$) & (M$_{\odot}$/yr) & (Gyr$^{-1}$)\\
\hline
40$^{+19}_{-8}$&68$^{+13.3}_{-23.6}$&10.47$^{+0.006}_{-0.08}$&4.27$^{+0. 7}_{-0.4}$&5.06$^{+0.07}_{-0.91}$& 20.2$^{+3.6}_{-0.3}$\\
\hline
\end{tabular}
\label{table:measurements}
\end{table}

\begin{figure}
\epsfig{file=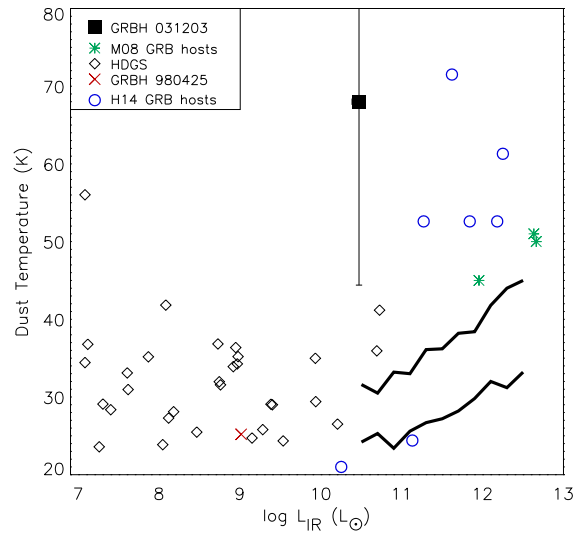, width=0.99\linewidth} 
\caption{Dust temperature as a function of total infrared
  luminosity. GRBH\,031203 is denoted by a closed square. The diamonds
  are the \textit{H}DGS from
  Ma13 and RR13 and the cross is GRBH\,980425 from M13. The asterisks and open circles are
  the high redshift IR-detected GRB
  hosts from M08 and H14 respectively. The lines are the 1$\sigma$ bounds of the
$L_{\rm IR}-T_{\rm dust}$ relation from Symeonidis et al. (2013) representative 
of the $z<$2 IR-luminous galaxy population. }
\label{fig:temp_lir}
\end{figure}

\section{The properties of GRBH\,031203}
\label{sec:properties}

In this section we examine the dust properties of GRBH\,031203 and
compare to the following sources: (i)
the \textit{Herschel} dwarf galaxy sample
(\textit{H}DGS; Madden et
al. 2013\nocite{Madden13}, hereafter Ma13 and R{\'e}my-Ruyer et al
2013\nocite{Remy-Ruyer13}, hereafter RR13), consisting of 48 dwarf galaxies at
$z<$0.05 and selected
to span the largest range in SFR and metallicity for dwarf galaxies in
the local Universe, (ii) GRBH\,980425 (Micha{\l}owski et al. 2013;
hereafter M13),
the only other GRB host galaxy in the local Universe observed by \textit{Herschel}
(iii) the 3 SCUBA-detected GRB hosts at $z\sim1$ from Micha{\l}owski et
al. (2008\nocite{Michalowski08}; hereafter M08) and (iv) the 7
\textit{Herschel}-detected GRB hosts at $0.3<z<2.5$ from Hunt et
al. (2014\nocite{Hunt14}; hereafter H14).

\begin{figure}
\epsfig{file=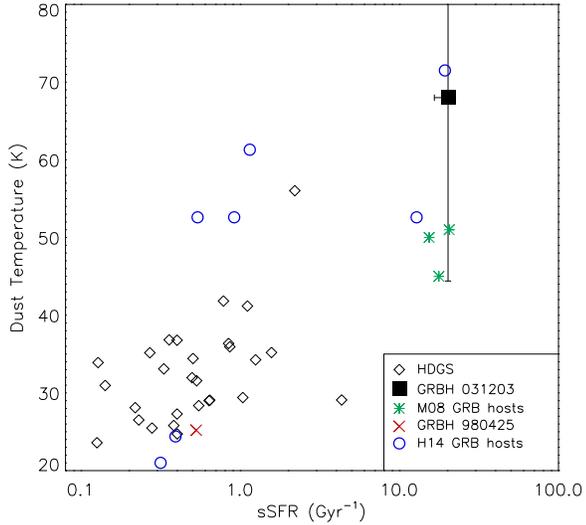, width=0.99\linewidth} 
\caption{Dust temperature as a function of $sSFR$. GRBH\,031203 is
  denoted by a closed square. The diamonds are the \textit{H}DGS from
  Ma13 and RR13 and the cross is GRBH\,980425
  from M13. The asterisks and open circles are
  the high redshift IR-detected GRB
  hosts from M08 and H14 respectively.}
\label{fig:temp_ssfr}
\end{figure}

Fig. \ref{fig:temp_lir} shows the locus of the above samples on the IR-luminosity - dust temperature ($L_{\rm IR}-T_{\rm dust}$) plane; for
reference we also plot the 1$\sigma$ bounds of the
$L_{\rm IR}-T_{\rm dust}$ relation from Symeonidis et al. (2013\nocite{Symeonidis13a}) representative 
of the $z<$2 IR-luminous galaxy population. 
Note that GRBH\,031203 has a warm dust temperature and an IR luminosity in the regime of
IR-luminous galaxies ($L_{\rm
  IR}>10^{10}$\,L$_{\odot}$). Interestingly, within the 1$\sigma$
uncertainty, its dust temperature is closer to the one typically seen
in many of the high redshift IR-detected M08 and H14 GRB hosts, than to the \textit{H}DGS and GRBH\,980425.
Fig. \ref{fig:temp_ssfr} shows $T_{\rm dust}$ versus $sSFR$; dust temperature is seen to
correlate better with $sSFR$ than $SFR$ in IR-luminous galaxies (e.g. Magnelli et al. 2013\nocite{Magnelli13}), suggesting that the
intensity of star-forming activity, rather than the star-formation rate, has a direct effect on the average
dust temperature. Again the locus of GRBH\,031203 is within the
parameter space probed by many of the high-redshift IR-detected GRB
hosts. In
contrast, GRBH\,980425 has a much smaller $sSFR$ and a lower
dust temperature.

\begin{figure}
\epsfig{file=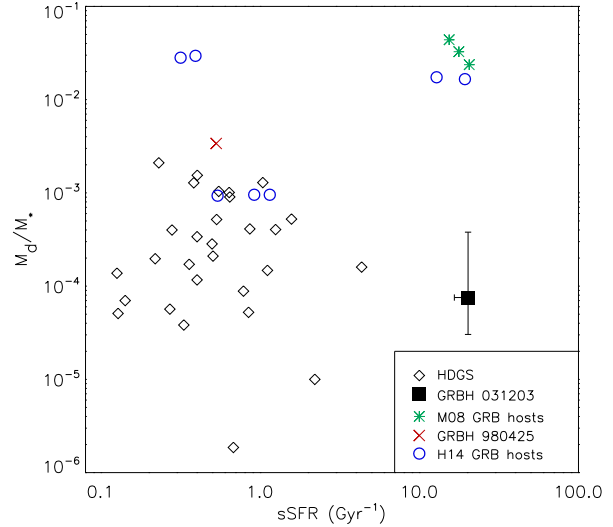, width=0.99\linewidth} 
\caption{The dust-to-stellar mass ratio as a function of specific
  star-formation rate. GRBH\,031203 is denoted by a closed square. The
  diamonds are the \textit{H}DGS from
  Ma13 and RR13 and the cross is GRBH\,980425 from M13. The asterisks and open circles are the
  high redshift IR-detected GRB
  hosts from M08 and H14 respectively.}
\label{fig:dust_sfr}
\end{figure}

\begin{figure}
\epsfig{file=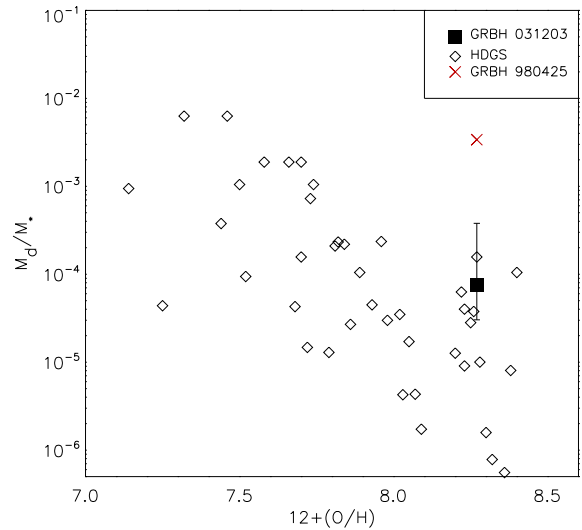, width=0.99\linewidth} 
\caption{The dust-to-stellar mass ratio as a function of
  metallicity. GRBH\,031203 is denoted by a closed square. The
  diamonds are the \textit{H}DGS from
  Ma13 and RR13, the cross is GRBH\,980425 from M13.}
\label{fig:metallicity_dust}
\end{figure}

Figs \ref{fig:dust_sfr} and \ref{fig:metallicity_dust} 
show the dust-to-stellar mass ratio as a function
of specific star formation rate and metallicity respectively. The
\textit{H}DGS spans a large range in dust-to-stellar mass ratio
($M_{d}/M_{\ast}$), however all values lie at $\lesssim$0.002, in agreement with normal star-forming
galaxies (SFGs) locally (e.g. Skibba et al. 2011\nocite{Skibba11}). This is in contrast to the high-redshift IR-detected GRB
hosts whose values are higher, in the 0.001$<M_{d}/M_{\ast}<$0.1 range, consistent with the
general IR-luminous galaxy population (e.g. Santini et
al. 2010\nocite{Santini10}). Interestingly, $M_{d}/M_{\ast}$ for
GRBH\,031203 is well within the range of the \textit{H}DGS, but
significantly lower than GRBH\,980425 and the $z\sim1$ IR-detected GRB
hosts.

\section{Summary and Conclusions}
\label{sec:conclusions}

We have reported \textit{Herschel}/PACS observations of the host
galaxy of GRB\,031203. Using the PACS data and ancillary IR
photometry, we have, for the first time, been able to place constraints
on the IR SED shape, total
infrared luminosity, IR-derived star formation rate, dust mass and dust
temperature of GRBH\,031203. We compared our findings with a
representative sample of local dwarf galaxies, high redshift IR-detected GRB hosts and the nearby well studied GRB host 980425, the
only other GRB host at $z<0.1$ to have \textit{Herschel}
observations. 

We found that GRBH\,031203 has a warm average dust temperature, a high specific
star-formation rate and an IR luminosity placing it in the regime of
IR-luminous galaxies. Interestingly, these properties are comparable
to those of the high-redshift IR-detected GRB
host galaxies and unlike what is seen in local dwarfs. On the other
hand its value of $M_{d}/M_{\ast}$ is within the range probed by local
dwarf galaxies. GRBH\,031203 is overall more active than typical local galaxies within
the same metallicity and stellar mass range, consistent with previous
reports of higher $sSFR$ amongst GRB hosts
compared to other SFGs (e.g. Castro Cer{\'o}n et
al. 2006\nocite{Castro-Ceron06}; Savaglio et
al. 2009\nocite{SGLB09}; Perley et al. 2013). Its large specific
star-formation rate indicates a starburst episode in action, in agreement with other
studies of GRBH\,031203 which conclude that it is a young system
undergoing a starburst, the supporting evidence according to Watson et
al. (2011) being the values of emission line
ratios (e.g. [NeIII]/[NeII]), lack of polycyclic aromatic
hydrocarbon (PAH) features and a low value of the 4000$\AA$ break
(D4000$<$1). 

Interestingly, GRBH\,031203 is also
in a separate class to the well studied GRBH\,980425. Although
GRBH\,031203 has similar metallicity to
GRBH\,980425, their $T_{\rm dust}$, $sSFR$ and $M_{d}/M_{\ast}$ values
are distinctly different. GRBH\,031203 is a much more active system, whereas
GRBH\,980425 seems more quiescent, akin to local dwarfs. This is
interesting because it suggests that GRB host galaxies can indeed span a
large range in properties, making LGRBs an ideal tool in selecting
relatively unbiased samples of star-forming galaxies up to high redshift ($z<10$;
e.g. Cucchiara et
al. 2011\nocite{Cucchiara11}).

\section*{Acknowledgments}
This paper uses data from \textit{Herschel}'s photometer
PACS, developed by a consortium of institutes led by MPE (Germany) and including UVIE (Austria); KU Leuven, CSL, IMEC (Belgium);
CEA, LAM (France); MPIA (Germany); INAF-IFSI/OAA/OAP/OAT, LENS,
SISSA (Italy); IAC (Spain) and supported by the funding
agencies BMVIT (Austria), ESA-PRODEX (Belgium), CEA/CNES (France),
DLR (Germany), ASI/INAF (Italy), and CICYT/MCYT (Spain). RS is
supported by a Royal Society Dorothy Hodgkin Fellowship. NS is the recipient of an Australian Research Council Future Fellowship.

\bibliographystyle{mn2e}
\bibliography{references}

\label{lastpage}

\end{document}